\documentclass[namedreferences]{solarphysics}
\usepackage[optionalrh]{spr-sola-addons} 
\usepackage{graphicx}        
\usepackage{color}           
\usepackage{url}             

\begin{document}

\begin{article}

\begin{opening}

\title{Visibility of Prominences using the  He\,{\sc i} D$_3$ Line Filter on PROBA-3/ASPIICS Coronagraph\\ {\it 
Solar Physics}}

\author{S.~\surname{Jej\v ci\v c}$^{1,2}$\sep
        P.~\surname{Heinzel}$^{2}$\sep
        N.~\surname{Labrosse}$^{3}$\sep
        A.~N.~\surname{Zhukov}$^{4,5}$\sep  
	A.~\surname{Bemporad}$^{6}$\sep
        S.~\surname{Fineschi}$^{6}$\sep
        S.~\surname{Gun\'ar}$^{2}$\sep
	 }
\runningauthor{Jej\v ci\v c et al.}
\runningtitle{Narrow-Band He\,{\sc i} D$_3$ Line Filter on ASPIICS Coronagraph}

 \institute{$^{1}$  University of Ljubljana, Faculty of Mathematics and Physics, Ljubljana, Slovenia
                     email: \url{sonja.jejcic@guest.arnes.si} \\  
            $^{2}$ Astronomical Institute of the Czech Academy of Sciences, Ond\v{r}ejov, Czech Republic
                      email: \url{pheinzel@asu.cas.cz}, \url{stanislav.gunar@asu.cas.cz}\\
            $^{3}$ SUPA School of Physics \& Astronomy, University of Glasgow, UK 
		      email: \url{nicolas.labrosse@glasgow.ac.uk} \\
            $^{4}$ Solar-Terrestrial Centre of Excellence-SIDC, Royal Observatory of Belgium, Brussels, Belgium
		      email: \url{andrei.zhukov@sidc.be} \\ 
            $^{5}$ Skobeltsyn Institute of Nuclear Physics, Moscow State University, Moscow, Russia \\ 
 	    $^{6}$ INAF-Turin Astrophysical Observatory, Pino Torinese (TO), Italy
		      email: \url{bemporad1@oato.inaf.it}, \url{fineschi@oato.inaf.it} \\
            }  

\begin{abstract}

We determine an optimal width and shape of the narrow-band filter centered around the  He\,{\sc i} D$_{3}$ 
line for prominence and coronal mass ejection (CME) observations with the ASPIICS ({\it Association of Spacecraft for 
Polarimetric and Imaging Investigation of the Corona of the Sun}) coronagraph onboard the PROBA-3 ({\it Project 
for On-board Autonomy}) satellite, to be launched in 2020. We analyze  He\,{\sc i} D$_{3}$ line 
intensities for three representative non-LTE prominence models at temperatures 8, 30 and 100~kK computed by the 
radiative transfer code and the prominence visible-light (VL) emission due to Thomson scattering on the prominence 
electrons. We compute various useful relations at prominence line-of-sight (LOS) velocities of 0, 100, and 
300~km~s$^{-1}$ for 
20~\AA~wide flat filter and three Gaussian filters with full width at half maximum (FWHM) equal to 5, 10, and 
20~\AA~to show the relative brightness contribution of the  He\,{\sc i} D$_{3}$ line and the prominence VL to the 
visibility in a given narrow-band filter. 
We also discuss possible signal contamination by Na\,{\sc i} D$_{1}$ and D$_{2}$ lines which otherwise may be 
useful to detect comets. Results mainly show: i) an optimal narrow-band filter should be flat or somewhere 
between flat and Gaussian with FWHM of 20~\AA~in order to detect fast moving prominence structures, 
ii) the maximum emission in the He\,{\sc i} D$_3$ line is at 30~kK and the minimal at 100~kK, and
iii) the ratio of emission in the He\,{\sc i} D$_3$ line to the VL emission can provide a useful diagnostic for 
the temperature of prominence structures. This ratio is up to 10 for hot prominence structures, 
up to 100 for cool structures and up to 1000 for warm structures.

\end{abstract}


\end{opening}

\section{Introduction}
     \label{s-int} 

Coronagraphic observations of the Sun from space were so far limited so that only the outer
corona was detected. The instruments like  {\it Large Angle and Spectrometric Coronagraph} (LASCO) or {\it Ultraviolet Coronagraph Spectrometer} (UVCS) 
on board of the {\it Solar and Heliospheric Observatory} (SOHO) \cite{Bru95,Koh95}, or COR2 coronagraphs on the twin {\it Solar Terrestrial Relations 
Observatory} (STEREO) satellites \cite{How08} used external occulters placed in front of the telescopes, which resulted in occultation significantly
exceeding the solar disk radius. This was dictated by the amount of stray light. Note that the stray light is high in internally occulted coronagraphs 
like STEREO/COR1. However, by placing the external occulting disk further from the telescope itself, one can gradually reduce the stray light,
reach lower coronal altitudes, and thus study the innermost corona. This kind of experiment is designed
for the upcoming mission {\it Project for On-board Autonomy} (PROBA-3) of the European Space Agency (ESA). PROBA-3 is a technology-demonstration
mission within the ESA’s General Support Technology Programme (GSTP), aimed at testing the so-called ‘formation flight’ of two
satellites (with the expected launch in 2020). The two satellites, separated on their orbits by about 150 meters, will carry a large
externally occulted coronagraph, where the occulter will be placed on the front satellite oriented towards
the Sun and the telescope will be on the rear one. This coronagraph, called ASPIICS ({\it Association of Spacecraft for Polarimetric and Imaging 
Investigation of the Corona of the Sun}) \cite{Lam10,Ren16} will then be precisely oriented in space in order to observe the Sun and such kind of 
orientation will fulfill part of the formation-flight tests. With this configuration, the inner corona down to 1.08 solar radii 
({\it i.e.} about 60~Mm above the limb), can be observed with unprecedented quality of imaging. In a sense, this will simulate total solar eclipses,
with the advantage of getting extended time series of coronal images which will allow us, for the first
time, to study evolution and dynamics of the inner corona on time scales reaching six hours.

One of the science objectives is also observation of eruptive prominences and coronal mass ejections (CMEs) in the field of
view (FOV) of ASPIICS, using filter imaging techniques. The filter wheel is still under the design study and it is the main objective of this
paper to propose its transmittance properties. We use our previous experience with the He\,{\sc i} D$_3$ line emission
in cool prominence-like structures \cite{Lab15} and study their expected visibility through the ASPIICS coronagraph.
This visibility critically depends on the physical conditions in eruptive prominences and cores
of CMEs, as well as on their velocities in the corona, and depends on the transmittance profile of the
narrow-band He\,{\sc i} D$_3$ filter. We also discuss a potential contamination of the signal by prominence
emission in nearby sodium doublet Na{\,\sc i} D$_1$ and D$_2$. The point is that if the filter will be wide
enough, the latter lines can be well detected in comets passing through the ASPIICS FOV. 
Images acquired by space-based coronagraphs have proven to be very useful for observations of near-Sun comets and even sun-grazing \cite{Bie02}
and sun-skirting \cite{Lam13} comets. These comets have been observed in the  visible-light (VL) even by the STEREO coronagraphs \cite{Tho09}, and a few of them have 
been observed also in UV spectra \cite{Bem07} and EUV images \cite{Mcc13}. All these observations provided lot of information about the origin 
and evolution of these small bodies in our solar system \cite{Sek03}.

While this paper is aimed at proposing the narrow-band filter transmittance characteristics for 
He\,{\sc i} D$_3$ line, in the
following paper we will focus on the diagnostic potential of this He\,{\sc i} D$_{3}$ line, combined with VL, 
and that will serve for future analysis of the ASPIICS data. In fact, there already exists the broad-band data of this kind for
structures at higher altitudes. LASCO-C2 coronagraph on SOHO has several broad-band filters, one
of them centered around the He\,{\sc i} D$_3$ lines (the so-called orange filter). Several observations
of prominence-like structures with this orange filter exist in the literature \cite{bou97} and
the question thus arises whether the image is dominated by the He\,{\sc i} D$_3$  line or by VL integrated through the broad-band
filter. The latter is due to Thomson scattering on electrons which are part of the erupting structure
(the coronal component subtracted). 
 The optimum parameters of the He\,{\sc i} D$_3$ filter, contamination with sodium D$_1$ and D$_2$ lines and comet detection capability with ASPIICS coronagraph are the main objectives of this article.

\section{Narrow-Band Filter Observations of Neutral Helium He\,{\sc i} D$_{3}$ Line}
	\label{s-nbf}

The ASPIICS coronagraph onboard the PROBA-3 satellite will carry a filter wheel with total of six filter slots: broad-band VL filter between 5400 and 5700~\AA, three VL polarizers at angles 0, and $\pm$ 60$^\circ$, iron Fe\,{\sc xiv} green-line narrow-band 5~\AA~ filter and finally one narrow-band filter to detect plasmas in  
the neutral helium  He\,{\sc i} D$_{3}$ at 5877.25~\AA~(we use the vacuum wavelength). 
The design of the He\,{\sc i} D$_{3}$ narrow-band filter is still in progress. 

The narrow-band filter centered around the He\,{\sc i} D$_{3}$ line contains the radiation contribution from four sources:
i) prominence emission in the He\,{\sc i} D$_{3}$ line mainly due to scattering of the photospheric radiation on prominence helium atoms, ii) VL 
emission due to Thomson scattering on prominence electrons, iii) VL emission from the surrounding corona along the line-of-sight (LOS) due to
Thomson scattering on coronal electrons, and iv) coronal He\,{\sc i} D$_{3}$ line emission. Here we do not consider iii) because the LOS contribution of 
the surrounding corona can be subtracted using the nearby pre-eruption coronal observation and iv) which is below the detection limit due to coronal 
temperatures much higher than the formation temperature of the He\,{\sc i} D$_{3}$ line.

We focus on three types of Gaussian filters with FWHM equal to 5, 10, and 20~\AA, to study the relative brightness contribution of  
the He\,{\sc i} D$_{3}$ line and the prominence VL to the visibility in a given filter. We also take into account the possible 
contamination with nearby sodium Na\,{\sc i} D$_{1}$ and D$_{2}$ lines at 5897.56 and 5891.58~\AA. 
The advantage of ASPIICS coronagraph on PROBA-3 is to have nearly simultaneous narrow-band He\,{\sc i} D$_{3}$ line and broad-band VL 
observations for the whole FOV down to 1.08~$R_\odot$ with strongly reduced stray light. But since ASPIICS  will not provide 
spectroscopic He\,{\sc i} D$_{3}$ observation, there will be no information about the prominence LOS velocity. Without such information we can not 
quantitatively interpret the prominence He\,{\sc i} D$_{3}$ emission unless the He\,{\sc i} D$_{3}$ narrow-band filter has flat response function in the 
wavelength range corresponding to expected Doppler shifts. In the present work we compare the effect of the prominence LOS velocity for three Gaussian 
filters with different FWHM, against the reference 20~\AA~wide flat filter.

\section{Modeling of the Prominence Emission in He\,{\sc i} D$_3$ Line}
	\label{s-ed3}

Emergent intensities in the He\,{\sc i} D$_{3}$ line are computed using the non-LTE ({\it i.e.} departures from Local Thermodynamic Equilibrium) 
radiative-transfer code presented in \inlinecite{Lab01} for a grid of prominence models. 
The code solves the pressure-balance and ionisation 
equilibrium equations, as well as radiative transfer and statistical equilibrium equations for hydrogen in a one-dimensional (1D) plane-parallel slab
vertically standing above the solar surface. 
Once these equations are solved, radiative transfer and statistical equilibrium equations are solved for a 34 level helium atom model, with 29 levels for 
He\,{\sc i}, four levels for He\,{\sc ii}, and one continuum level for fully ionised helium. This allows us to obtain the emergent intensities as a function of 
model input parameters. The input parameters are temperature, gas pressure, microturbulent velocity, radial flow velocity, slab thickness, and altitude 
above the limb. The slab is externally illuminated by the solar disk.
For this exploratory study, we focus on isothermal and isobaric prominence slabs at three altitudes above the limb 
(60 Mm, 800~Mm, and 1600~Mm, roughly corresponding to 1.08~$R_\odot$, 2.15~$R_\odot$, and 3~$R_\odot$) representative of what could be 
observed with the ASPIICS coronagraph. We computed a total of 90 models with temperatures ranging from 8~kK to 100~kK, thickness of 1000~km and 
5000~km, and microturbulent velocities of 5~km s$^{-1}$, 15~km s$^{-1}$, and 20~km s$^{-1}$ ({\it i.e.} increasing with temperature). These models 
are static because there is no Doppler Brightening Effect (DBE) in the He\,{\sc i} D$_{3}$ line (the photospheric spectrum around the He\,{\sc i} D$_{3}$ line 
is flat {\it i.e.} continuum) and we neglect the velocity effects on the electron density.
DBE is discussed in \inlinecite{Hei87}. The helium abundance is fixed at 0.1. The non-LTE code provides the total energy emitted in the helium D$_{3}$ 
line $E_{\rm D3}$, the maximal intensity of the He\,{\sc i} D$_{3}$ profile $I_{\rm 0}$, the mean electron density $n_{\rm e}$ as well as the optical 
thickness at the line center of the He\,{\sc i} D$_{3}$ line $\tau_{0}(\rm D_{3})$. Note that the radiative transfer codes have been used for the 
first time to model hot prominence structures inside the CMEs.

\begin{table}  
\caption{Temperature $T$, microturbulent velocity $\xi$, mean electron density $n_{\rm e}$, integrated intensity 
emitted in the He\,{\sc i} D$_{3}$ line $E_{\rm D3}$, optical thickness at the line center of the 
He\,{\sc i} D$_{3}$ line $\tau_{0}(\rm D_{3})$, and optical thickness of VL 
$\tau_{\rm VL}$ for three representative non-LTE models. Models have fixed gas pressure $p$ = 0.1~dyn~cm$^{-2}$, 
thickness $D$ = 1000~km, and height above the solar surface $h$ = 60~Mm.}
\label{t-ms}
\begin{tabular}{ccccccc} 
 \hline model~&{\it T}~&{\it $\xi$}~&{\it n$_{\rm e}$}~&{\it E$_{\rm D3}$}~& {\it $\tau_{0}(\rm D_{3})$}~&{\it $\tau_{\rm VL}$} \\
           &(kK)~&(km~s$^{-1}$)~&(10$^{10}$~cm$^{-3}$)~&(erg~s$^{-1}$~cm$^{-2}$~sr$^{-1}$)&(10$^{-3})$~&(10$^{-7})$\\
 \hline  1&8&5&2.16&657.5&2.40&14.4\\
         2&30&15&1.26&2237.9&3.50&8.4\\
         3&100&20&0.378&8.4&0.01&2.5\\
 \hline
\end{tabular}
\end{table} 

Our initial analysis revealed that the He\,{\sc i} D$_{3}$ line is formed mainly under two different regimes, namely due to scattering of the incident 
radiation coming from the disk (for models at low temperatures and low pressures), or collisional excitation (for models with large temperatures or large 
pressures). This is in line with the analysis of \inlinecite{Lab04}. In order to determine the optimal filter around the 
He\,{\sc i} D$_{3}$ line for the ASPIICS coronagraph, it is sufficient to select only a subset of our grid of 90 models.
Therefore, we present detailed results only for three temperature models: 8~kK, 30~kK, and 100~kK. These correspond to three cases 
of prominence ejecta which can be observed by the ASPIICS coronagraph -- cool,  warm, or hot prominence material in the corona.
Note that the latter one was identified as a prominence flux-rope in the core of the CME emitting in hydrogen Lyman lines and other hotter lines 
\cite{Hei16,Jej17}. At 8~kK, the He\,{\sc i} D$_{3}$ line formation is dominated by scattering of the incident radiation in the He\,{\sc i} D$_{3}$ line. At 
the two higher temperatures, 
collisional processes will dominate. In particular, all models at 30~kK show maximum emission in the He\,{\sc i} D$_{3}$ line: at  temperatures higher than 
30~kK, neutral helium starts to be noticeably ionized, and the intensity of the emergent He\,{\sc i} D$_{3}$ line decreases with temperature. 
For this study, all other input parameters are fixed to one set values: altitude above the limb of 60 Mm, gas pressure of 0.1~dyn cm$^{-2}$, slab 
thickness of 1000~km. The microturbulent velocity is taken to be 5~km s$^{-1}$ at $T$ = 8~kK, 15~km s$^{-1}$ at $T$ = 30~kK, and 20~km s$^{-1}$ 
at $T$ = 100~kK (see Table~\ref{t-ms}). A more detailed analysis of the He\,{\sc i} D$_{3}$ line intensity and how it will be observed in the 
ASPIICS filter will be presented in a follow-up paper \cite{Lab18}.

\section{Visible-Light Emission}
	\label{s-evl}

The integrated VL intensity of a prominence $E_{\rm VL}$ is due to Thomson scattering of the solar incident radiation on free electrons 
in the prominence plasma and depends on $n_{\rm e}$ and $D$
\begin{equation}
E_{\rm VL} = \sigma_{\rm T}~W(h, \lambda_{\rm c})~I_{\rm tot}~n_{\rm e}~D  \ .
\label{e-evl}
\end{equation}
\begin{table}  
\caption{Computed values of VL intensities $I_{\rm tot}$ and integrated VL emissions $E_{\rm VL}$ for different 
filters and for three representative prominence models.}
\label{t-evl}
\begin{tabular}{cccccc} 
 \hline Filter~&Flat~&~&Gaussian~&~&Broad-band \\
	FWHM~(\AA)~&20~&5~&10~&20~&300 \\
 \hline {\it I$_{\rm tot}$~}~(10$^{7}$~erg~s$^{-1}$~cm$^{-2}$~sr$^{-1}$)&6.46&1.72&3.43&6.87&101.44 \\
       {\it E$_{\rm VL}$~}(\#1)~(erg~s$^{-1}$~cm$^{-2}$~sr$^{-1}$)&21.76&5.79&11.58&23.15&335.90 \\
         {\it E$_{\rm VL}$}(\#2)~(erg~s$^{-1}$~cm$^{-2}$~sr$^{-1}$)&12.69&3.38&6.75&13.51&195.94 \\
	 {\it E$_{\rm VL}$}(\#3)~(erg~s$^{-1}$~cm$^{-2}$~sr$^{-1}$)&3.81&1.01&2.03&4.05&58.78 \\
 \hline
\end{tabular}
\end{table} 
VL emission is thus proportional to the column electron density $N_{\rm e} = n_{\rm e}~D$. Here $\sigma_{\rm T}$ is the Thomson scattering cross section.
Note that the optical thickness is $\tau_{\rm VL} = \sigma_{\rm T}~n_{\rm e}~D$ (see Table~\ref{t-ms}). 
$W(h, \lambda_{\rm c})$ is the dilution factor which depends on height above the solar surface and on the central wavelength of the filter due to the limb darkening of incident radiation. 
\citeauthor{Jej09} (\citeyear{Jej09}; see Appendix therein) show the height and wavelength dependence of the dilution factor in the optical range up to the altitude 100~Mm.
Note that here we are interested in VL emission around the He\,{\sc i} D$_{3}$ line and thus the wavelength of the He\,{\sc i} D$_{3}$ line center is used.
$I_{\rm tot}$ (see Table~\ref{t-evl}) has to be computed in the wavelength range in which the VL is detected
\begin{equation}
I_{\rm tot} =~\int~I_{0}(\lambda)~G(\lambda)~{\rm d}\lambda \ .
\label{e-itot}
\end{equation}
Here $I_{0}(\lambda)$ is the VL continuum photospheric intensity at a given wavelength using the data from \inlinecite{All73} and $G(\lambda)$ is the 
Gaussian-shaped filter transmittance around the He\,{\sc i} D$_{3}$ line with FWHM equal to 5, 10, and 20~\AA~for narrow-band filters, or uniform 
transmittance for 300~\AA~broad-band VL filter between 5400 and 5700~\AA~and a narrow-band flat filter with the width of 20~\AA~(see 
Section~\ref{s-nbf}). 
For three representative models discussed in Section~\ref{s-ed3}, VL emission is computed using Equation~\ref{e-evl} and the results 
are presented in Table~{\ref{t-evl}} for all studied filters.

ASPIICS coronagraph will detect the mixture of the He\,{\sc i} D$_{3}$ and VL contributions through the narrow-band filter 
\begin{equation}
E_{\rm tot}({\rm nb}) = E_{\rm D3}({\rm nb}) + E_{\rm VL}({\rm nb})  \ .
\label{e-mix}
\end{equation}
Here we use (nb)  for a narrow-band filter to clearly distinguish from the broad-band (bb) filter.
Since the instrument will make  quasi-simultaneous images in the broad-band VL filter, reconstruction of the pure helium image and pure VL 
image of the prominence can be done. VL emission in the broad-band filter can be written in the same way as in narrow-band one
(see Equations~\ref{e-evl} and \ref{e-itot}). 
If broad-band and narrow-band intensities are calibrated in absolute units one can get
\begin{equation}
E_{\rm VL}({\rm nb}) = E_{\rm VL}({\rm bb})~\frac{W(h, \lambda_{\rm nb})}{W(h, \lambda_{\rm bb})}~\frac{\int~I_{0}(\lambda)~G_{\rm nb}(\lambda)~{\rm d}\lambda }{\int~I_{0}(\lambda)~
G_{\rm bb}(\lambda)~{\rm d}\lambda }= E_{\rm VL}({\rm bb})~\alpha \ ,
\label{e-evlbb}
\end{equation}
where $\alpha$ is a known parameter for a given height through the dilusion factors which depend on the filter (for nb we take wavelength of the He\,{\sc i} D$_{3}$ line~and for bb we take 5550~\AA) and filter transmittance shape. 
Pure He\,{\sc i} D$_{3}$ contribution in the narrow-band filter can be then written as
\begin{equation}
E_{\rm D3}(\rm nb) = E_{\rm tot}({\rm nb}) - E_{\rm VL}({\rm nb}) \ .
\label{e-ped3}
\end{equation}

\section{Visibility of the He\,{\sc i} D$_3$ Prominence Structures with ASPIICS Coronagraph}
	\label{s-vis}

\begin{figure}    
\centerline{\includegraphics[width=0.75\textwidth,clip=]{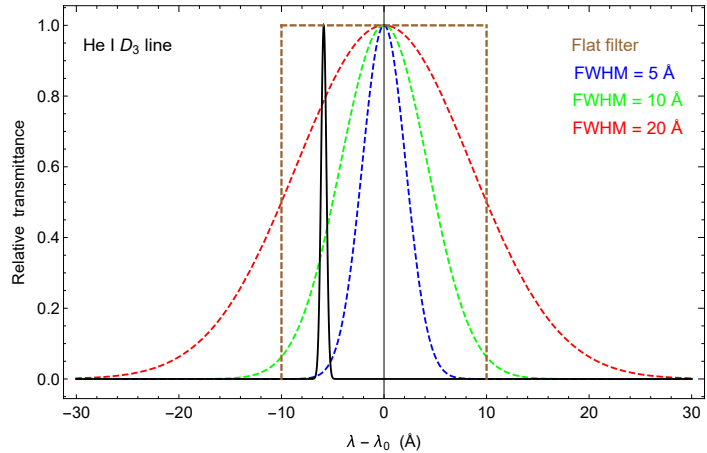}
            }
\caption{Transmittance normalized to the peak values as a function of wavelength around the He\,{\sc i} D$_{3}$ 
line for all four studied narrow-band filters together with the Doppler shifted He\,{\sc i} D$_{3}$ profile at LOS velocity 300~km~s$^{-1}$.}
\label{f-fs}
\end{figure}

Figure~\ref{f-fs} shows the effect of the LOS velocity 300~km~s$^{-1}$ on the prominence He\,{\sc i} D$_{3}$ line due to the filter 
transmission curves at four different narrow-band filters. 
Narrow-band 20~\AA~flat filter is marked in brown and three Gaussian filters have the following color coding: blue for 5~\AA, green for 10~\AA, 
and red for 20~\AA. This color coding is used for all following plots. Note that for the He\,{\sc i} D$_{3}$ line the Doppler shift of 
1~\AA~corresponds roughly to the LOS velocity 50~km~s$^{-1}$. This plot clearly shows that at large Doppler shifts of the prominence structure only 
the flat filter will transmit the whole signal of the He\,{\sc i} D$_{3}$ line while Gaussian filters will reduce signal by a certain factor. 
At this rather extreme LOS velocity the helium signal would be reduced by about 20 \% in case of the 20~\AA~filter, by about 60  \% in case of the 10~\AA~ filter and it is practically completely reduced by the 5~\AA~filter. 
This demonstrates that the optimal choice would be the 20~\AA~flat narrow-band filter. Possible option is a design of the 20~\AA~narrow-band filter with 
transmittance between flat and Gaussian shape.  
 
\begin{figure}    
\centerline{\hspace*{0.015\textwidth}
            \includegraphics[width=0.9\textwidth, clip=]{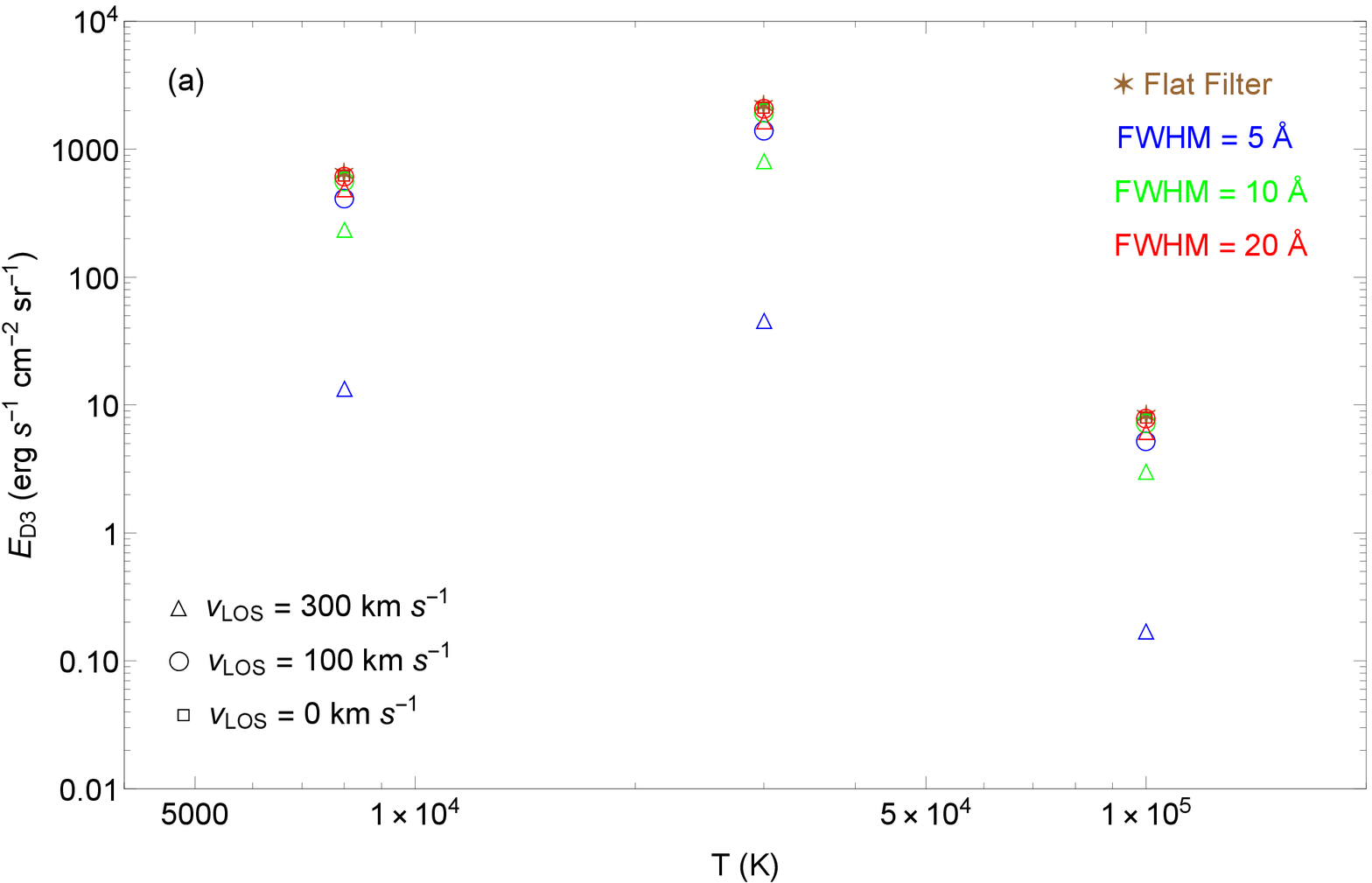}
            \hspace*{-0.02\textwidth}
            }
\vspace{0.01\textwidth}
\centerline{\hspace*{0.015\textwidth}
            \includegraphics[width=0.9\textwidth,clip=]{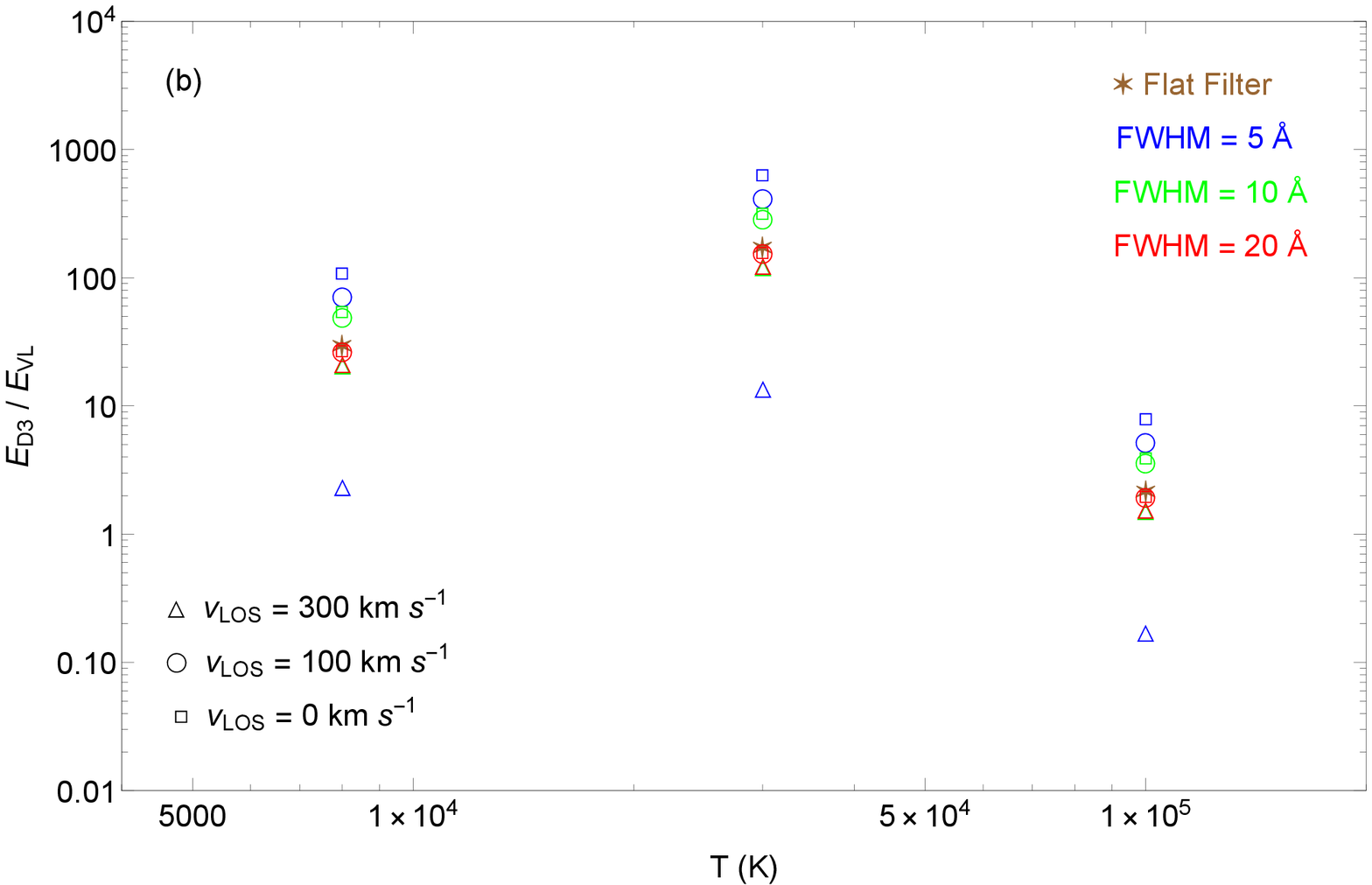}
            \hspace*{-0.03\textwidth}
            }
\caption{a) Integrated intensity emitted in the He\,{\sc i} D$_{3}$ line  as a function of temperature for the studied models shown in 
Table~\ref{t-ms}. 
b) Ratio of the integrated intensity emitted in the He\,{\sc i} D$_{3}$ line to the VL emission as a function of temperature for the 
studied models.}
\label{f-ms}  
\end{figure}

\begin{figure}    
\centerline{\includegraphics[width=0.9\textwidth,clip=]{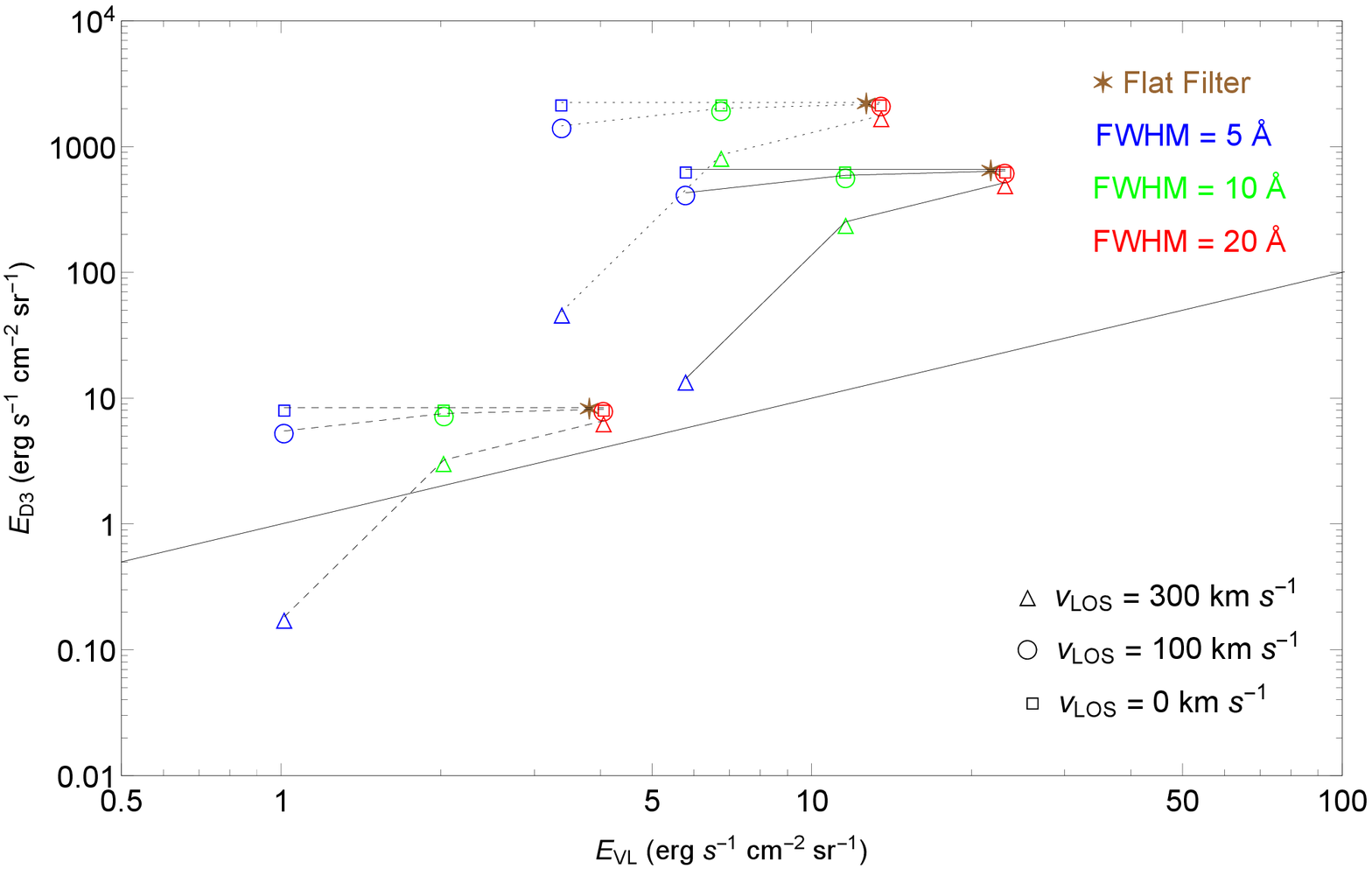}
            }
\caption{Integrated intensity of the He\,{\sc i} D$_{3}$ line as a function of VL emission for selected models. Model 1 is presented by 
{\it solid line}, model 2 by {\it dotted line} and model 3 by {\it dashed line}. We show the effect of various filters and the prominence 
LOS velocities. {\it Straight black solid line} shows where $E_{\rm D3} = E_{\rm VL}$.}
\label{f-sp}
\end{figure}

Figure~\ref{f-ms}a shows the variation of the integrated intensity emitted in the He\,{\sc i} D$_{3}$ line with temperature for four narrow-band filters and 
three LOS velocities, for three prominence models shown in Table~\ref{t-ms}. 
Note that we choose two representative LOS velocities 100, and 300~km~s$^{-1}$ according to our previous experiences with fast moving prominence structures \cite{Hei16}, and we add a zero velocity as the reference for static prominence structures.
$E_{\rm D3}$ is the highest for warm prominence structures at 30~kK and the lowest for hot prominence model at 100~kK where the helium is almost fully ionized. 
Flat filter is a reference filter which gives the correct energy emitted in the He\,{\sc i} D$_{3}$ line because the filter transmittance is uniform. 
By increasing the prominence LOS velocity from 0 up to 300~km~s$^{-1}$, the energy detected in the 
He\,{\sc i} D$_{3}$ line is decreasing for narrower 
Gaussian filters. Difference between flat and Gaussian filter is highest at 5~\AA~and lowest at 20~\AA. 

\begin{table}  
\caption{Ratio of integrated intensity emitted in the He\,{\sc i} D$_{3}$ line to the prominence VL emission for different prominence LOS velocities and models. 
Note that for Gaussian filters FWHMs are presented.}
\label{t-rms}
\begin{tabular}{ccccc} 
\hline Filter  & Flat~20~\AA~ & Gaussian~5~\AA~& Gaussian~10~\AA~& Gaussian~20~\AA \\
\hline
{$v_{\rm LOS}$}~(km~s$^{-1}$)~&~&~&model 1~&\\
\hline
   0  &    &    113.57   &    56.80   &    28.40 \\
   100  &   30.22&   74.22  &   51.07  &   27.65 \\
   300  &  &   2.47  &   21.81  &   22.36 \\
\hline
{$v_{\rm LOS}$}~(km~s$^{-1}$)~&~&~&model 2~&\\
\hline
   0  &    &    662.66   &    331.43   &    165.70 \\
   100  &   176.31&   433.04  &   297.99 &   161.36\\
   300  &  &   14.40  &  127.26  &   130.44 \\
\hline
{$v_{\rm LOS}$}~(km~s$^{-1}$)~&~&~&model 3~&\\
\hline
   0  &    &    8.29  &   4.15   &   2.07 \\
   100  &  2.21&   5.42  &  3.73 &   2.02\\
   300  &  &  0.18  &  1.59  &   1.63 \\
\hline
\end{tabular}
\end{table}

By separating the integrated intensity of the He\,{\sc i} D$_{3}$ line and prominence VL emission using the broad-band filter as described in 
Section~\ref{s-evl}, we can reconstruct similar plot as in Figure~\ref{f-ms}a. Figure~\ref{f-ms}b shows the ratio of integrated intensity emitted in 
the  He\,{\sc i} D$_{3}$ line to the prominence VL emission as a function of temperature for representative models. Ratios are presented in 
Table~\ref{t-rms}. Because both the He\,{\sc i} D$_{3}$ and the narrow-band VL emissions are optically thin (see Table~\ref{t-ms}), their ratio is independent 
on the prominence thickness meaning that it may provide a useful diagnostic of the temperature. 
For a given range of prominence LOS velocities and narrow-band filters the ratio of energy emitted 
in the  He\,{\sc i} D$_{3}$ line to VL emission is between 0.1 and 10 for hot prominence structures, between 1 and 100 for cool structures, and between 10 and 1000 for warm structures.
Figure~\ref{f-sp} shows the effect of various filters and the prominence LOS velocities on the integrated intensity of the He\,{\sc i} D$_{3}$ line and the 
VL emission. The plot clearly shows that for lower LOS velocities 0, and 100~km~s$^{-1}$ the contrast of the He\,{\sc i} D$_{3}$ line with respect to the prominence VL 
emission is decreasing with the filter width, while for larger LOS velocities 300~km~s$^{-1}$ the contrast is inceasing with the filter width (see connecting lines). 
In order to be able to detect fast moving structures, we must find a compromise between the filter which will compose the broadest range of LOS velocities 
and the width providing the best detectibility of the He\,{\sc i} D$_{3}$ emission. In Figure~\ref{f-sp} the majority of points lie above the diagonal which 
indicates a good detectability of the He\,{\sc i} D$_{3}$ line emission. For the range of prominence LOS velocities up to 
300~km~s$^{-1}$, the Gaussian filter with 20~\AA~FWHM gives the closest values to the flat filter which indicates that a real design of the 
narrow-band filter centered around the He\,{\sc i} D$_{3}$ line should be the 20~\AA~flat filter or somewhere between flat and Gaussian filter with FWHM of 20~\AA.

\section{Contamination with Na\,{\sc i} D$_1$ and D$_2$ Lines and Comet Detection Capability of ASPIICS}
	\label{s-d12}

Sodium doublet Na\,{\sc i} D$_{1}$ and  D$_{2}$ at 5897.56 and 5891.58~\AA~lies close to the 
He\,{\sc i} D${_3}$ line and may contaminate the signal 
for fast moving prominence structures. 
If the narrow-band filter will be wide enough, sodium lines can be well detected in comets passing through the ASPIICS FOV, which represents an interesting option for ASPIICS.

The magnitude of comets observed in VL  coronagraphs is not only due to photospheric light scattered by dust grains, but also to emission processes by 
neutrals or singly ionized atoms. In particular, for many comets a distinct neutral sodium tail has been reported \cite{Bie02}. Strong Na\,{\sc i} 
emission at small heliocentric distances have been detected in many comets 
({\it e.g.} comet Ikeya-Seki, \citeauthor{Pre67} \citeyear{Pre67}, \citeauthor{Sla69} \citeyear{Sla69}), 
and a distinct Na\,{\sc i} tail was first imaged by \inlinecite{Cre97} at comet Hale-Bopp (C/1995 O1). This emission is very bright thanks to the high 
efficiency of sodium D transition (see {\it e.g.} \citeauthor{Cre99} \citeyear{Cre99} for a discussion of sodium tail origin). 
The sodium tail is in addition to the two most common dust and ion tails, and is responsible for 
significant differences in the apparent comet magnitudes depending on the band-pass filter used for coronagraphic observations, as discussed by 
\inlinecite{Kni10}. 
This brightness increase (by about a factor 10) is relatively more important for "narrow-band" (such as the "orange" LASCO-C2 and LASCO-C3 filters, 
5400 - 6400 \AA) with respect to broad-band (such as the "clear" LASCO-C3 filter, 4000 - 8500 \AA) filters, and is most likely due to the sodium 
D$_{1}$ and D$_{2}$ lines, even if contributions from other lines by other atoms or ions cannot be excluded. Significant differences have been 
observed for instance between light curves for the same 
comet (ISON C/2012 S1) observed with LASCO and STEREO coronagraphs \cite{Kni14}, and these differences are related mostly to the fact that the bandpasses
for COR1 (6500 - 6600 \AA) and COR2 (6500 - 7500 \AA) coronagraphs do not include the sodium lines. 

As reported by \inlinecite{Bie02}, the majority of sun-grazing comets usually disappear below heliocentric distances of about 7~$R_{\odot}$, hence no 
emission from these comets is expected in the ASPIICS FOV. Nevertheless, recent spectacular observations of near Sun comet (like ISON C/2012 S1) and 
sun-grazing comets (like Lovejoy C/2011 W3) demonstrate that similar objects could fall in the ASPIICS FOV, and even survive at their perihelion transit. 
In particular, during the transit of comet ISON it was demonstrated for the first time that similar comets can be observed in VL coronagraphs imaging the 
inner corona even from the ground \cite{Dru14}. Hence, the inclusion of the D$_{1}$ and D$_{2}$ lines in the future ASPIICS He\,{\sc i} D$_{3}$ band-pass 
filter centered at 5877.25~\AA~could allow similar observations for future comets. Nevertheless, this would imply broad band-passes up to 28~\AA~ and 
40~\AA~wide to include, respectively, the sodium D$_{1}$ and D$_{2}$ lines.

Effects due to the cometary orbital motion should be considered as well. The cometary emission from sodium atoms will be in general Doppler shifted, 
if the comet will have a significant projected velocity along the LOS, that could be in general directed both towards the observer or away from it. 
The maximum possible speed for these bodies is equal to the free-fall speed that, by assuming for  instance an heliocentric distance of perihelion of 
1.2~$R_{\odot}$, is about 560 km~s$^{-1}$, corresponding to a maximum expected Doppler shift by about 11~\AA. The sodium atoms expelled from the comet 
will undergo radiation pressure acceleration in the anti-sunward direction, hence these atoms will likely move at a speed lower than the cometary orbital 
speed. In any case, by assuming a He\,{\sc i} band-pass filter centered at 5877.25~\AA~ with a full width by 20~\AA, the cometary sodium emission at 5891.58 
and 5897.56~\AA~ will start to fall in the band-pass only for Doppler blue-shifts larger than 4~\AA~ and 10~\AA, corresponding to velocities larger than 
200 km~s$^{-1}$ and 500 km~s$^{-1}$.
In summary, cometary sodium emission will likely fall out of the narrow-band He\,{\sc i} D$_{3}$ filter, unless much broader band-passes are chosen. 
This will significantly reduce the detection probability of comets with this filter, because sodium lines will fall in the filter band-pass only for 
significant Doppler blueshifts, that are possible but quite unlikely. Nevertheless, bright comets will be observed as well by ASPIICS with the VL filters 
(5400 - 5700 ~\AA), providing nice imaging of the dust tails in the inner corona, and could be even observed in the "green line" filter, as recently 
demonstrated by the AIA observations of comet Lovejoy  \cite{Mcc13}. Hence, by considering that the He\,{\sc i} D$_{3}$  filter is designed primarily to 
perform scientific observations of solar prominences, this probable loss of detection capability of comets can be considered as acceptable.

\begin{figure}    
\centerline{ \vspace{0.005\textwidth}
\includegraphics[width=0.48\textwidth,clip=]{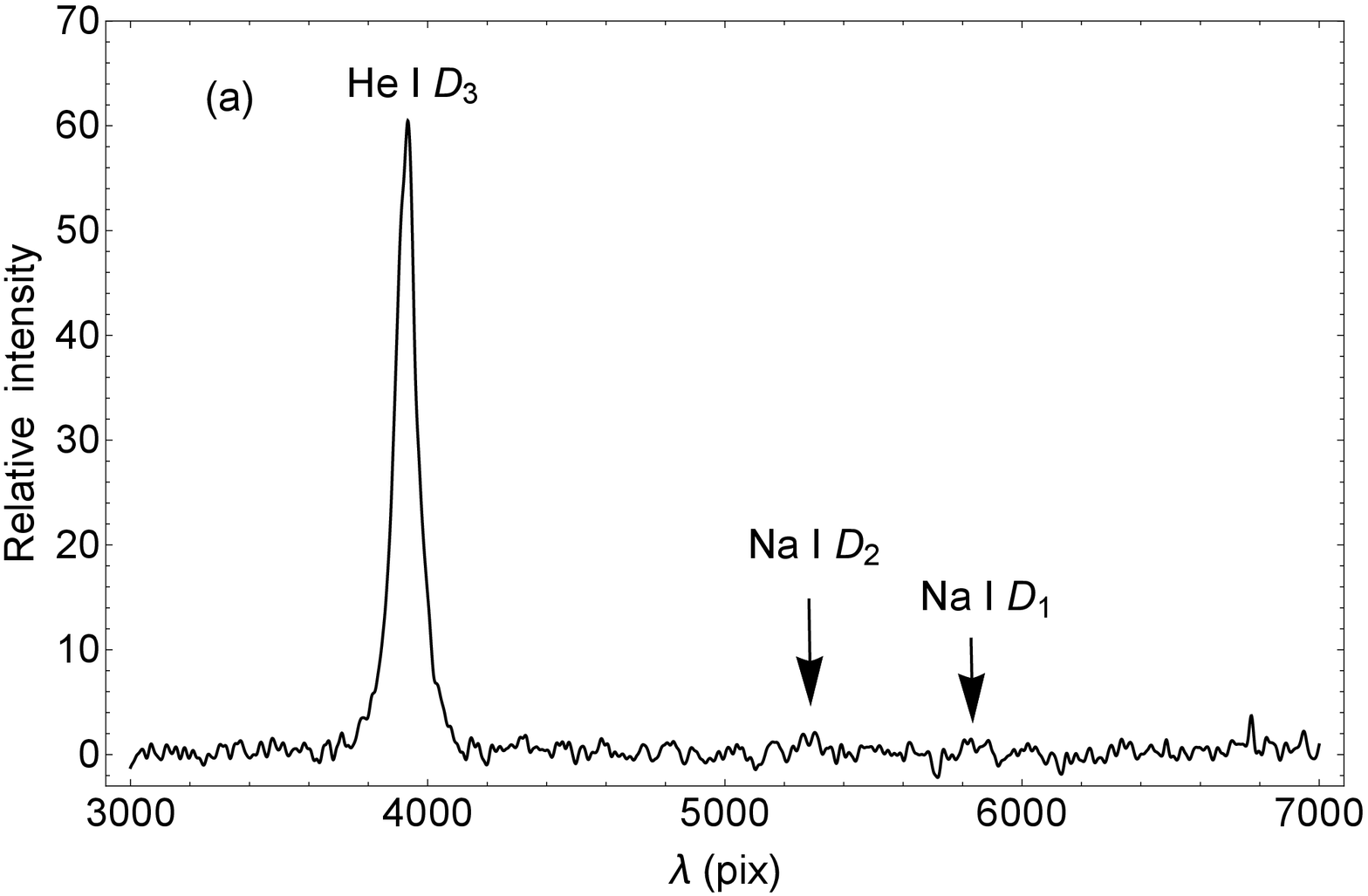}
            \includegraphics[width=0.48\textwidth,clip=]{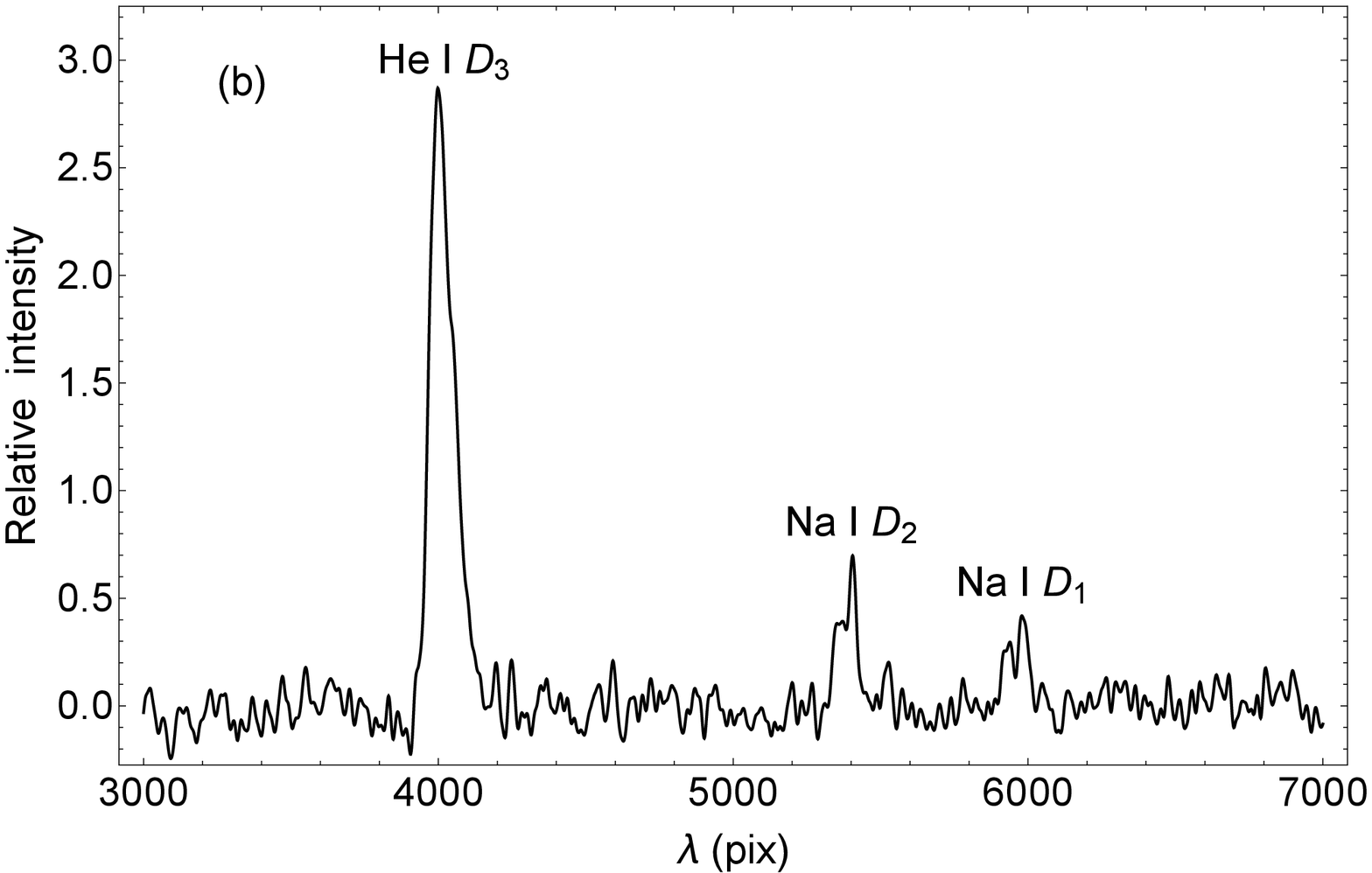}
            }
\caption{a) Example of measured profiles of the He\,{\sc i} D$_{3}$ and the Na\,{\sc i} D$_{1}$ and D$_{2}$ lines. Contribution of Na\,{\sc i} D$_{1}$ and D$_{2}$ 
lines is negligible. b) Contribution of Na\,{\sc i} D$_{1}$ and D$_{2}$ lines is between 10 and 15 \% with respect to the He\,{\sc i} D$_{3}$ line.}
\label{f-nap}
\end{figure}

To study the possible contribution of sodium Na\,{\sc i} D$_{1}$ and D$_{2}$ lines in prominences observed by narrow-band He\,{\sc i} D$_{3}$ filter, we use the spectra 
obtained in the past on photographic plates at the Ond\v{r}ejov Observatory. These spectra contain many prominences observed in the He\,{\sc i} D$_{3}$ and 
the sodium D$_{1}$ and D$_{2}$ lines. Photographic densities were converted to relative intensities with standard 
procedure using the calibration wedge \cite{Val59}. The photospheric scattered light was subtracted. An example where the contribution of sodium  D$_{1}$ and D$_{2}$ lines in prominences 
is negligible compared to the 
He\,{\sc i} D$_{3}$ line is shown in Figure~\ref{f-nap}a, while Figure~\ref{f-nap}b 
shows an example where contribution of  sodium D$_{1}$ and D$_{2}$ lines in the prominence is important.

\begin{figure}    
\centerline{\includegraphics[width=0.475\textwidth,clip=]{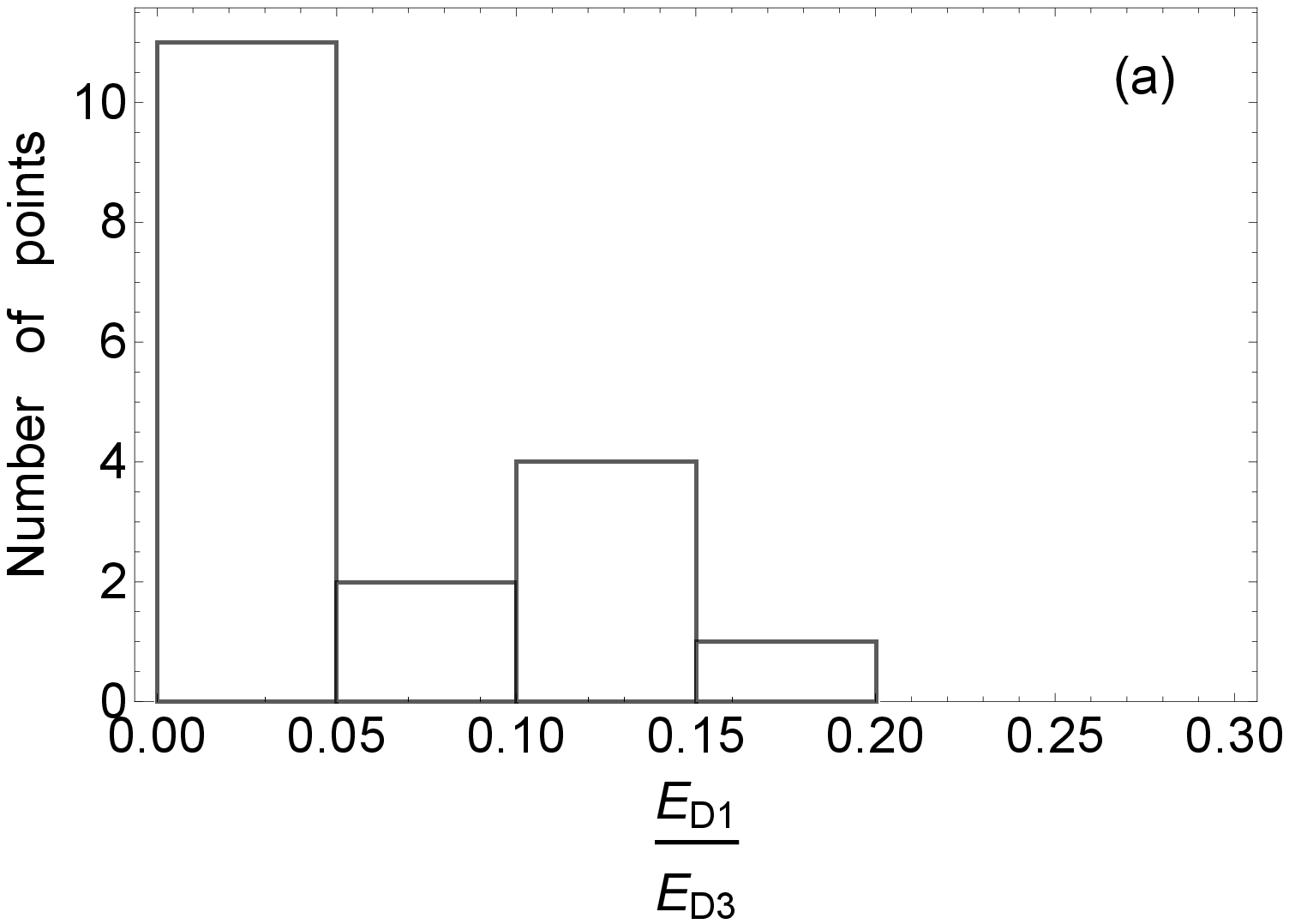}
            \hspace*{0.0\textwidth}
            \includegraphics[width=0.475\textwidth,clip=]{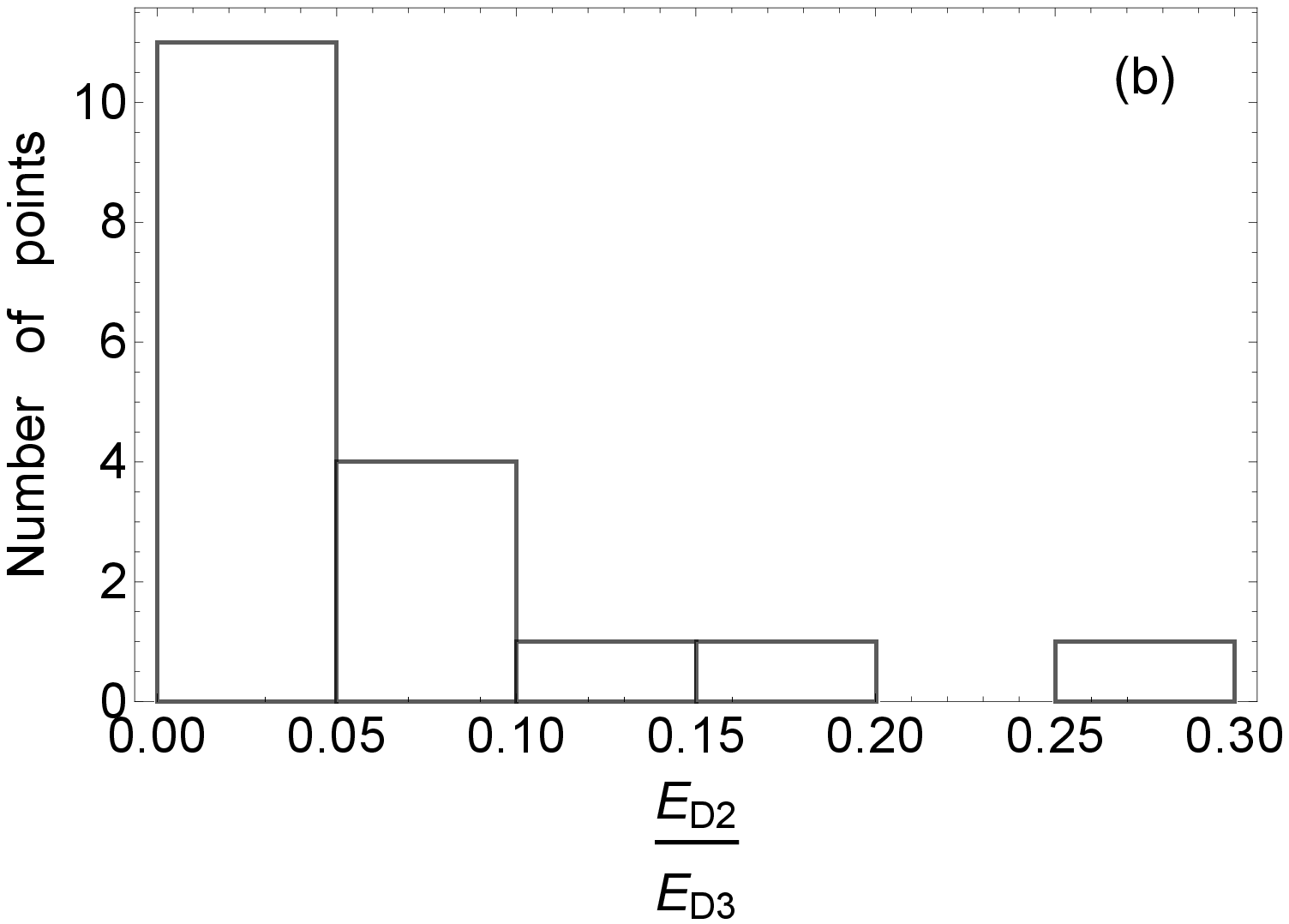}
            }
\caption{a) Distribution of ratio of energy emitted in Na\,{\sc i} D${_1}$ line to the He\,{\sc i} D$_{3}$ line. b) Distribution of  ratio of energy emitted in 
Na\,{\sc i} D${_2}$ to the He\,{\sc i} D$_{3}$ line.}
\label{f-his}
\end{figure}

In total we analyzed 18 prominence spectra from the Ond\v{r}ejov Observatory. Results are presented as histograms in Figure~\ref{f-his}
showing relative brightness contribution of Na\,{\sc i} D$_{1}$ and Na\,{\sc i} D$_{2}$ lines versus 
the He\,{\sc i} D$_{3}$ line. Histograms show that peak is below 5~\%. These are mostly quiescent prominences, in bright quiescent prominences the ratio 
can go up to 10 \%. These results are consistent with the work of \inlinecite{Ste05} who got similar results for the Na\,{\sc i} D$_{2}$ line. Only in a 
few cases ratio between Na\,{\sc i} D$_{1}$ and He\,{\sc i} D$_{3}$ or Na\,{\sc i} D$_{2}$ and He\,{\sc i} D$_{3}$ can be larger than 20 \%. 
These extreme cases seem to be related with the 
DBE in  the sodium D$_{1}$ and D$_{2}$ lines. Note that  the sodium doublet is  very sensitive to DBE, because 
these photospheric absorption lines are very deep and relatively narrow.  

\begin{table}  
\caption{Total energy emitted in Na\,{\sc i} D$_{1}$ and D$_{2}$ lines with respect to energy emitted in the He\,{\sc i} D$_{3}$ line at upper limit of radial flow 
velocity 450 km~s$^{-1}$ and for various LOS velocities at blueshift and redshift.}
\label{t-db}
\begin{tabular}{ccccc}
 \hline FWHM~&{\it v$_{\rm rad}$}~&{\it v$_{\rm LOS}$}~&$\frac{E_{\rm D2}+E_{\rm D1}}{E_{\rm D3}}$~&
          $\frac{E_{\rm D2}+E_{\rm D1}}{E_{\rm D3}}$~ \\
                                                       \\
          (\AA)~&~(km~s$^{-1}$)&~(km~s$^{-1}$)&Blueshift~&Redshift\\
 \hline  &  & 0 & 0.036 & 0.036\\
         20 & 450 & 100 & 0.055 & 0.024 \\
         &  & 300 & 0.128 & 0.010\\
 \hline  &  & 0 & 0.095 & 0.095\\
         30 & 450 & 100 & 0.116 & 0.078 \\
         &  & 300 & 0.172 & 0.053\\
 \hline
\end{tabular}
\end{table} 

\begin{figure}    
\centerline{\includegraphics[width=0.49\textwidth,clip=]{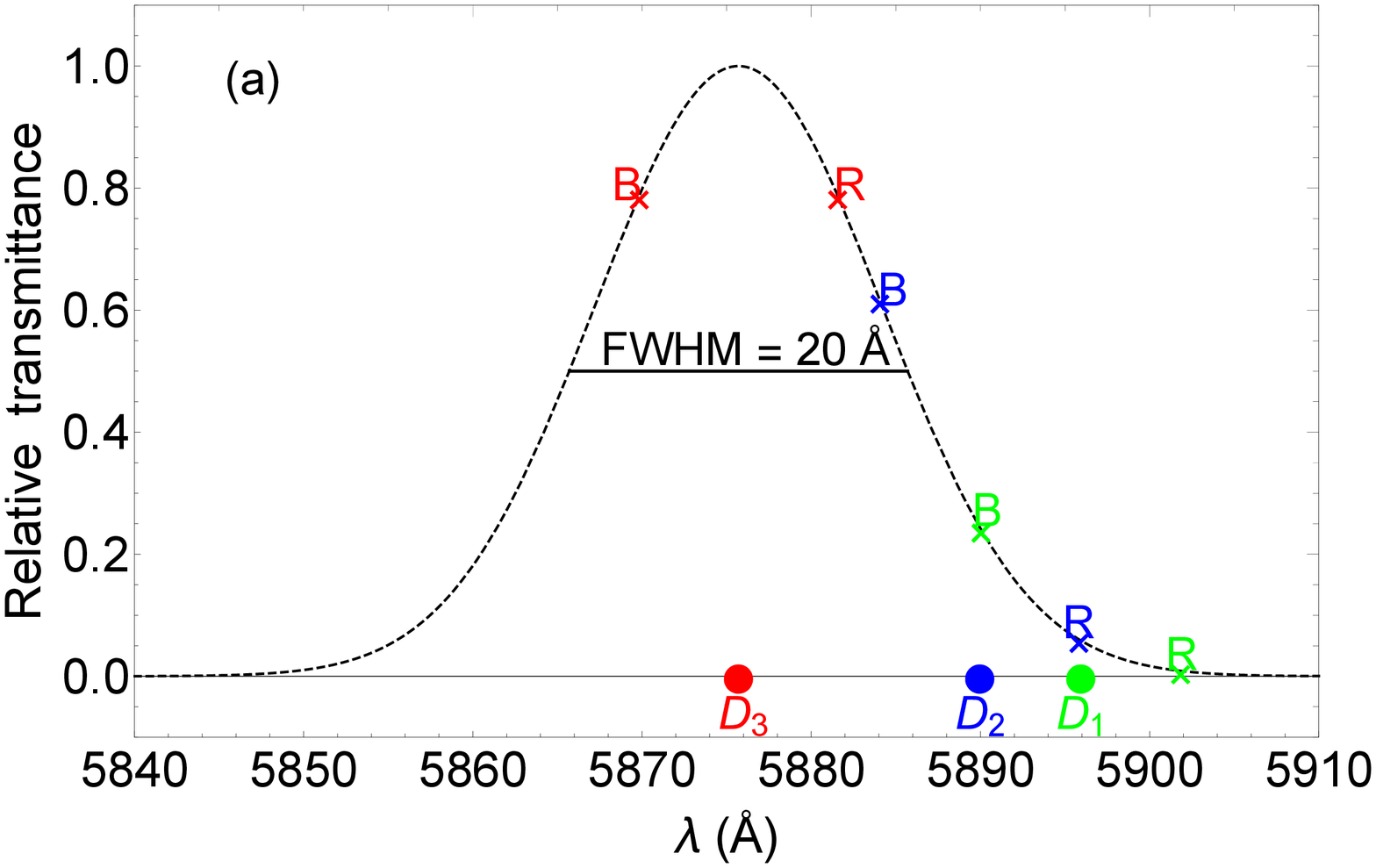}
            \hspace*{0.0\textwidth}
            \includegraphics[width=0.49\textwidth,clip=]{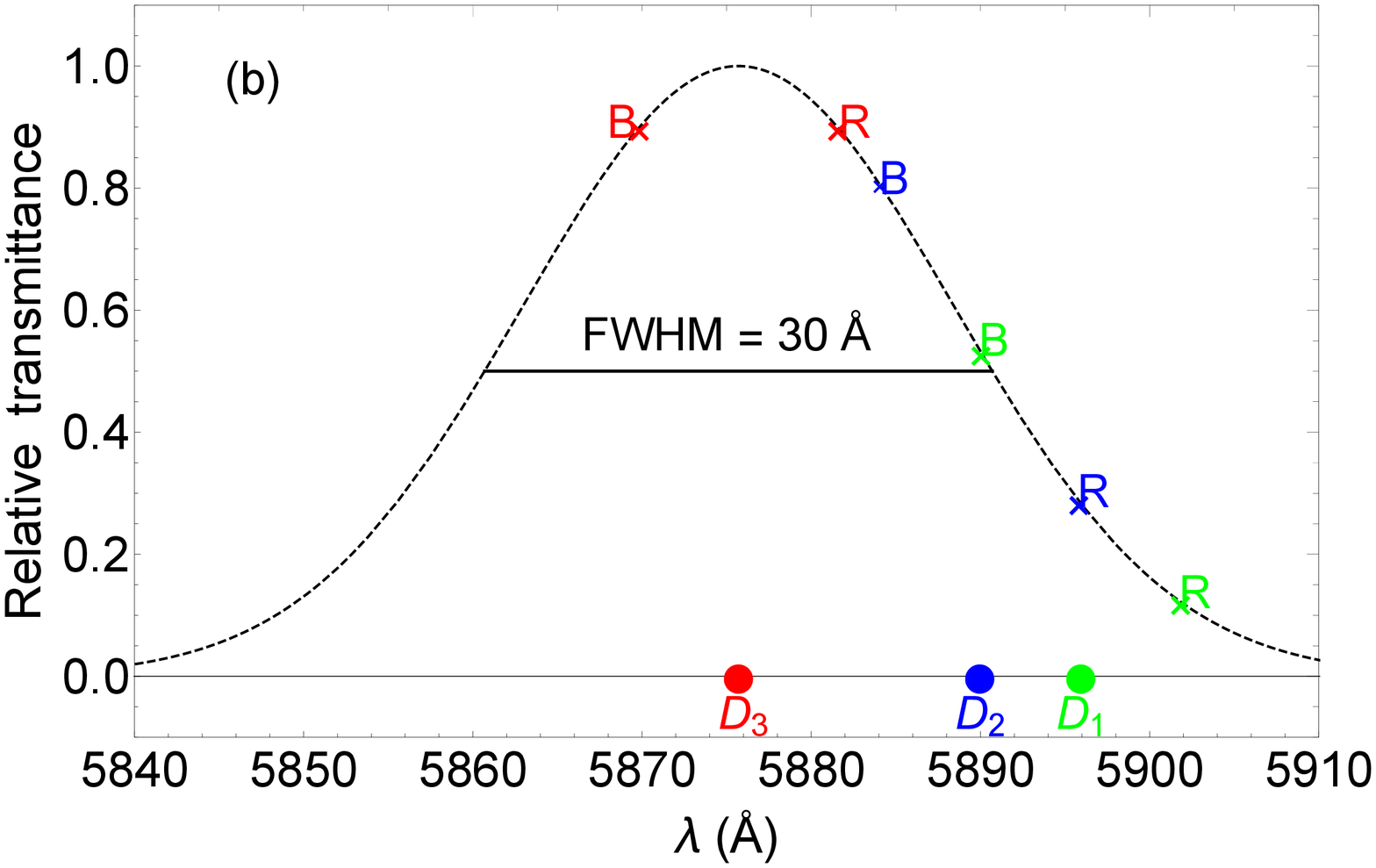}
            }
\caption{a) Contamination of Gaussian 20~\AA~narrow-band filter with the Na\,{\sc i} D$_{1}$ and D$_{2}$ lines at LOS velocity $\pm$ 300 km~s$^{-1}$. 
He\,{\sc i} D$_{3}$ line is marked in red, Na\,{\sc i} D$_{1}$ and D$_{2}$ lines in green and blue. Crosses show the position of all three lines at given LOS velocities. 
$R$ stands for redshift and $B$ for blueshift. b) Contamination of Gaussian 30~\AA~narrow-band filter with the  
Na\,{\sc i} D$_{1}$ and D$_{2}$ lines at 
LOS velocities $\pm$ 300 km~s$^{-1}$.}
\label{f-nac}
\end{figure}

It is important to know how much each line contributes to the narrow-band filter signal to see how important are 
Na\,{\sc i} D$_{1}$ and D$_{2}$ lines 
compared to the He\,{\sc i} D$_{3}$ line and how their ratio depends on the width of the filter. For that we made two examples with Gaussian narrow-band filter with FWHM equal to 20 and 30~\AA. 
In Table~\ref{t-db} we computed the relative contribution of Na\,{\sc i} D$_{1}$ and D$_{2}$ lines and the He\,{\sc i} D$_{3}$ line to the filter signal at an 
upper limit of the radial flow velocity $v_{\rm rad}$ = 450 km~s$^{-1}$ (which takes into account DBE on 
the sodium D$_{1}$ and D$_{2}$ lines) for three 
different prominence LOS velocities 0, 100, and 300 km~s$^{-1}$ moving in two directions: toward the observer (blueshift) or away from the observer 
(redshift) for two filter widths. 
In the 20~\AA~filter the contributions of Na\,{\sc i} D$_{1}$ and D$_{2}$ lines and VL are almost negligible, a prominence will be well visible in 
He\,{\sc i} D$_{3}$, exception is only an extreme case with about 13 \% of Na\,{\sc i} D$_{1}$ and D$_{2}$ contamination. 
With the 20~\AA~wide He\,{\sc i} D$_{3}$ filter ASPIICS can detect comets in Na\,{\sc i} D$_{1}$ and D$_{2}$ lines only 
if they move with LOS velocity 300 km~s$^{-1}$ or higher toward the observer (see Figure~\ref{f-nac}a). 
In the 30~\AA~Gaussian filter the prominence VL emission gets brigher which would lower He\,{\sc i} D$_{3}$ line contrast. 
Note that at 30~\AA~wide filter VL emissions of models 1, 2 , and 3 are 34.68, 20.23, and 6.07 
erg~s$^{-1}$~cm$^{-2}$~sr$^{-1}$. If there is no LOS motion of the prominence, we will automatically 
get about 10 \% of emission from the Na\,{\sc i} D$_{1}$ and D$_{2}$ lines through the filter. At a LOS velocity 300 km~s$^{-1}$ (see Figure~\ref{f-nac}b), 
we get 17 \% of Na\,{\sc i} D$_{1}$ and D$_{2}$ lines 
at blueshift which is not negligible contribution and 5 \% at redshift.  In the 30~\AA~filter the prominence would be more
contaminated by Na\,{\sc i} D$_{1}$ and D$_{2}$ lines and prominence VL emission. However, for that filter comets will be better detected while prominences 
will be more contaminated with the Na\,{\sc i} D$_{1}$ and D$_{2}$ contribution. Our analysis therefore shows that a 20~\AA~filter will have an optimal 
detectability of the He\,{\sc i} D$_{3}$ line in eruptive prominences.

\section{Conclusions}
	\label{s-con}

In this study we investigated the properties of an optimal narrow-band filter centered around the neutral helium He\,{\sc i} D$_{3}$ line at 5877.25~\AA~for 
eruptive prominence and CME observations down to 1.08~$R_{\odot}$ (60~Mm), with the ASPIICS coronagraph onboard PROBA-3 satellite. 
We have studied the expected visibility of prominences through various narrow-band filters in the ASPIICS coronagraph. This depends on the filter 
transmittance profile and on physical conditions in prominences. 
We took into consideration four narrow-band filters: three Gaussian filters with FWHM equal to 5, 10, and 20 \AA~and one 20~\AA~flat filter. 
A narrow-band He\,{\sc i} D$_{3}$ filter contains the radiation contribution mainly from the prominence He\,{\sc i} D$_{3}$ 
line and prominence VL emission. The modeled prominence He\,{\sc i} D$_{3}$ line emission was computed by multilevel 
1D non-LTE code for helium and the emergent line profiles were multiplied by the narrow-band filter transmittance. 
The integrated VL emission in prominences was also computed for all four narrow-band filters.  
We presented several relations between the prominence He\,{\sc i} D$_{3}$ line emission and the prominence VL emission depending on three typical 
prominence  LOS (Doppler) velocities: 0, 100 and 300~km~s$^{-1}$ in a given narrow-band filter. 
We have selected three representative prominence models with temperatures 8~kK 
(cool prominence structures), 30~kK (warm structures), and 100~kK (hot structures). 
These relations show: i) the maximum emission in the He\,{\sc i} D$_3$ line at 30~kK and the minimal at 100~kK, 
ii) by increasing the prominence LOS velocity, $E_{\rm D3}$ is decreasing when narrowing the Gaussian filter, 
iii) the contrast of $E_{\rm D3}$ with respect to $E_{\rm VL}$ for lower LOS velocities 0 and 100~km~s$^{-1}$ is decreasing with increasing filter width, while at 300~km~s$^{-1}$ is increasing, and finally
iv) $E_{\rm D3}$/$E_{\rm VL}$ can provide a useful diagnostic for the temperature of prominence structures. Models show that this ratio (which depends on prominence LOS velocity and filter width)
is between 0.1 and 10 for hot prominence structures, 
between 1 and 100 for cool structures and between 10 and 1000 for warm structures.
Note that He\,{\sc i} D$_{3}$ filter will produce a mixture of 
He\,{\sc i} D$_{3}$ line emission and VL emission (at a given wavelength position of the filter transmittance profile), therefore we can reconstruct the 
pure helium image of the prominence, as well as pure VL image thanks to simoultaneous broad-band VL continuum observations, but without any information 
about the prominence LOS velocities. 
We also show possible contamination of the signal with the nearby sodium doublet Na\,{\sc i} D$_{1}$ and D$_{2}$ in 20 and 30~\AA~Gaussian filters. 
Our recomendation is to design a narrow-band 20~\AA~flat filter or a filter
somewhere between flat and Gaussian with FWHM of 20~\AA. This is mainly dictated by the lack of information about the prominence LOS velocity.
However, a 30~\AA~filter (flat or almost flat) would better suit expected  cometary detections in Na\,{\sc i} lines. Such filter will more degrade the 
He\,{\sc i} D$_{3}$ prominence contrast, but He\,{\sc i} D$_3$ will be still larger or comparable to VL emission. As we have shown, the prominence images in 
He\,{\sc i} D$_3$  and VL can be separated using the broad-band VL filter data. But 30~\AA~filter signal will be more contaminated by prominence sodium lines 
(Figure~\ref{f-nac}) and this contribution can not be separated. 20~\AA~filter centered off-band of He\,{\sc i} D$_{3}$ towards red was also an option, 
but then the blushifted prominences would be less detectable.
A more detailed analysis of the He\,{\sc i} D$_{3}$ line, its diagnostics and how it will be observed with the ASPIICS filter, will be presented in the 
next paper \cite{Lab18}.


\begin{acks}
SJ acknowledges the financial support from the Slovenian Research Agency No. P1-0188.
SJ, PH and SG acknowledge support from the Czech Funding Agency through the grant No. 16-18495S and from the AIAS 
through RVO-67985815. 
NL acknowledges support from STFC grant ST/L000741/1. 
NL, PH and SG also acknowledge support by the International Space Science Institute (ISSI), and namely through the 
Team of NL dealing with solar prominences. ANZ acknowledges support from the Belgian Federal Science Policy Office through the ESA-PRODEX programme (grant No. 4000117262). 
SG and PH acknowledge the support from grant 16-17586S of the Czech Science Fundation.
Ond\v{r}ejov archive spectra were digitized by Alicja Heinzel.
\end{acks}

\mbox{}~\\

\bibliographystyle{spr-mp-sola} 

\bibliography{biblio}  


\end{article} 

\end{document}